\documentclass[useAMS,usenatbib]{mn2e}
\usepackage{graphicx}
\usepackage{times}

\begin{document}

 \title[Turbulence model  with mid-IR images]{Testing turbulence model at metric scales 
with mid-infrared VISIR images at the VLT}

\author[A. Tokovinin, M. Sarazin, \& A. Smette]{A. Tokovinin$^{1}$\thanks{E-mail:
atokovinin@ctio.noao.edu}, M. Sarazin$^{2}$, and A.Smette$^{3}$ \\
$^{1}$Cerro-Tololo Inter American Observatory, Casilla 603, La Serena, Chile\\
$^{2}$European Southern Observatory,
Karl-Schwarzschild-Strasse, 2 D-85748
Garching bei M\"unchen, Germany  \\ 
$^{3}$European Southern Observatory, Alonso de Cordova 3107, Casilla 19001, Vitacura, Santiago, Chile  
}

\date{-}

\pagerange{\pageref{firstpage}--\pageref{lastpage}} \pubyear{2005}

\maketitle

\label{firstpage}

\begin{abstract}

We  probe turbulence structure  from centimetric  to metric  scales by
simultaneous imagery at mid-infrared and visible wavelengths at the VLT
telescope  and show that  it departs  significantly from  the commonly
used Kolmogorov  model. The data can  be fitted by  the von K\'arm\'an
model with an outer scale of the order of 30\,m and we see clear signs
of the  phase structure function saturation across  the 8-m VLT  aperture.  The
image  quality  improves in  the  infrared  faster  than the  standard
$\lambda^{-1/5}$  scaling  and   may  be  diffraction-limited  at  30-m
apertures even  without adaptive optics  at wavelengths longer  than 8
micron.
\end{abstract}


\section{Introduction}

As the ground-based telescopes become bigger, more emphasis is made on
studying and modeling atmospheric optical distortions at large spatial
scales. This knowledge is required,  e.g. for specifying the stroke of
deformable mirrors in adaptive  optics or the range of fringe-trackers
in interferometers.   Moreover, the size of  the atmospheric coherence
length  exceeds  1\,m  at   infrared  (IR)  wavelengths.   Thus,  even
classical  long-exposure  imagery  in   the  IR  is  affected  by  the
departures of turbulence statistics from the standard Kolmogorov model
(variance proportional to the 5/3  power of the baseline) which was so
successfully used in the visible range.

Optical  path-length  fluctuations  were measured  with  long-baseline
interferometers,      but      the      published     results      are
controversial. Saturation of the fringe motions at baselines from 8\,m
to   16\,m   was   first   observed  by   \citet{Mariotti84}.   Later,
\citet{Davis95} found a strong departure from the Kolmogorov law and a
saturation of the  path-length fluctuations at the 80-m  baseline at a
level of about 10\,$\mu$m  rms.  On the other hand, \citet{Colavita87}
have not found  any departure from the Kolmogorov  law at baselines up
to  12\,m by  doing  a temporal  analysis  of the  fringe motion,  and
\citet{Nightingale91} reached the same conclusion from interferometric
measurements of  the fringe motion  in a 4-m telescope  aperture.  The
controversy could be  caused by the use of  the frozen-flow hypothesis
in the interpretation  of fringe motion on a  single baseline. Another
reason is  the difficulty  in separating fringe  motion caused  by the
atmosphere from the mechanical  noise due to instrument instabilities,
tracking, etc.

Turbulence outer scale $L_0$ can  be evaluated from the covariances of
the  image motion  in  small  telescopes, as  implemented  in the  GSM
instrument \citep{GSM}, or from  the analysis of adaptive-optics data,
e.g.  by  \citet{Fusco04}.  These methods use the  von K\'arm\'an (VK)
turbulence  spectrum  (cf.   Appendix  A) and  adjust  its  parameters
$(r_0,L_0)$ to fit  the data.  Measurements with the  GSM at different
sites  show that  typical  $L_0$ values  are  of the  order of  20\,m.
\citet{Maire06}  compared directly  fringe motion  in  a long-baseline
interferometer  with  the $L_0$  measured  by  GSM  and found  a  good
agreement.

This short and non-exhaustive review demonstrates that further work on
characterizing large-scale  turbulence structure is  needed.  A direct
comparison of mid-IR and optical  images at a large telescope offers a
new, independent way to probe turbulence models at spatial scales from
centimeters to meters.  To our  knowledge, no such work has been done
previously, so we made an experiment at the Very Large Telescope (VLT)
located  at the  ESO  observatory  Cerro Paranal  in  Chile. The  main
purpose  of  this experiment  is  to  evaluate  the atmospheric  phase
structure function (SF) and to check  if the VK model is adequate.  So
far,  the mathematically  convenient VK  model has  been  used without
being actually tested.

The experiment is described in  Sect.~2.  The results are presented in
Sect.~3 and the conclusions  are given in Sect.~4.  Appendices contain
some technical  material.  For  reader's convenience we  reproduce the
formulation of atmospheric models in Appendix~A.

\section{Experiment description}

\begin{figure*}
\centerline{\includegraphics[width=14cm]{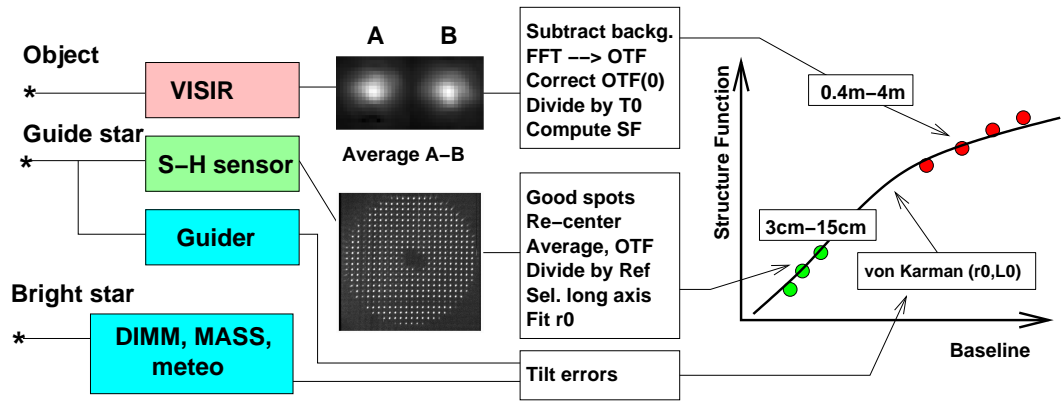} }
\caption{Overview of the experiment and data interpretation. 
\label{fig:exper} }
\end{figure*}

\subsection{From PSF to structure function}

It  is  well known  that  the  modulus  of the  long-exposure  optical
transfer function  (OTF) $T({\bf f})$ in  a perfect telescope is  related to
the   atmospheric  SF   $D_\varphi({\bf   r})$  as   \citep{Tatarsky61,
Roddier81}

\begin{equation}
T({\bf f}) = T_0({\bf f}) \exp [-0.5 D_\varphi(\lambda {\bf f}) ],
\label{eq:T}
\end{equation}
where $T_0({\bf f})$ is the  diffraction-limited OTF, ${\bf f}$ is the
vector of spatial frequency on the sky in rad$^{-1}$, $f = |{\bf f}|$,
$\lambda$  is the  imaging wavelength  and ${\bf  r}$ is  the baseline
vector.  This relation can  be inverted to reconstruct $D_\varphi({\bf
r})$ from the known image point spread function (PSF).  However, it is
only  feasible at  the baselines  ${\bf r}  = \lambda  {\bf  f}$ where
$D_\varphi({\bf  r})$ is not  very large  or small,  otherwise $T({\bf
f})$ is close to either 1  or 0 and the sensitivity to the atmospheric
turbulence  is  lost.   In   the  visible  range,  the  method  probes
centimetric and  decametric scales, in  the mid-IR it is  sensitive to
the metric scales because $\lambda$ is much larger.

The 8-m VLT telescope offers a unique platform for our experiment with
access  to metric baselines.   The aberrations  are removed  by active
optics and  the turbulence inside the  dome is low,  ensuring that the
image blur is dominated by  the atmospheric seeing.  Optical images as
small as $0.18''$ FWHM have been recorded under exceptional conditions,
proving that  the VLT  intrinsic quality is  nearly ideal even  in the
visible   range.\footnote{ESO   Press    Release,   July   21,   2000.
http://www.eso.org/outreach/press-rel/pr-2000/pr-16-00.html}  For  our
purpose, the mid-IR imager VISIR  installed at the Cassegrain focus of
the UT3 telescope is the best choice.

A deeper analysis of the mid-IR images shows that some departures from
the ideal-telescope model (\ref{eq:T}) are inevitable.  Residual image
motion (tilt)  is the largest source  of uncertainty, as  it can cause
additional blur (e.g.  wind shake),  while, on the other hand, part of
the atmospheric  tilt is removed by guiding.   Residual aberrations in
the VLT  optics and instrument can add something to the  atmospheric PSF,
too.     Therefore,   opto-mechanical   wave-front    distortions   of
instrumental nature  cannot be separated cleanly  from the large-scale
atmospheric  distortions. In this  respect our  new experiment  is not
fundamentally different from long-baseline interferometers, but it was
worth trying  nevertheless because instrumental effects  in both cases
are different.

\subsection{Overview}

\begin{table}
\center
\caption{Instrument parameters}
\label{tab:par}
\begin{tabular}{l l  l l}
\hline
Instrument &  VISIR  & VISIR & SH \\
Filter    &   PAH1  & Q2    & None \\   
\hline
Aperture diam., m & 8.115 & 8.115 & 0.34 square \\
$\lambda$/$\Delta \lambda$, $\mu$m & 8.6/0.42 & 18.7/0.88 & 0.6/$\sim$0.3 \\ 
Pixel scale, arcsec & 0.075 & 0.075 & 0.280 \\
Detector format &   256$\times$256 &  256$\times$256 &  592$\times$573 \\
Exposure time, s & 30$\times$2 &  90$\times$1 & $\sim$45 \\
Chopping period, s & 4 & 2 & None \\
\hline
\end{tabular}
\end{table}

The   overall    scheme   of   the   experiment    is   presented   in
Fig.~\ref{fig:exper}.   Two different stars  are observed  through the
VLT: the  object with  VISIR, the guide  star with  the Shack-Hartmann
(SH) sensor of the VLT active  optics (AO) system. The same guide star
is used for the guiding, called {\em field stabilization}. The angular
distance between the object and the  guide star was from $3'$ to $5'$.
Details of  the optical  and IR imagery  and data reduction  are given
below and in Table~\ref{tab:par}.

Data on the  seeing and turbulence profile at  the Paranal observatory
are  collected  by  the  dedicated  site  monitor  equipped  with  the
Differential   Image   Motion    Monitor   (DIMM)   \citep{DIMM}   and
Multi-Aperture Scintillation Sensor  (MASS) \citep{MASS}.  The monitor
points  to a bright  star near  zenith and  measures the  total seeing
$\epsilon$  and  the seeing  in  the  free atmosphere  $\epsilon_{FA}$
produced  by  turbulence  above  500\,m. Naturally,  $\epsilon_{FA}  <
\epsilon$  unless most  turbulence is  above 500\,m.  The VLT  dome is
higher than the DIMM tower, hence the seeing at VLT can be better than
that measured  by DIMM, but  still worse or equal  to $\epsilon_{FA}$.
The MASS also measures  the adaptive-optics time constant $\tau_0$ and
a  crude turbulence  profile.  The  effective wind  speed in  the free
atmosphere   $\overline{V}$  was   evaluated  from   the   relation  $
\overline{V} = 0.31  r_{0, FA}/\tau_0$.  The speed of  the ground wind
$V_{gr}$ was taken from the Paranal ambient conditions database.

\subsection{Conditions of the experiment}

\begin{table}
\center
\caption{Night log}
\label{tab:nights}
\medskip
\begin{tabular}{l cc cc c c  }
\hline
\noalign{\medskip}
Date & Time & Air  & $\epsilon$, & $\epsilon_{FA}$, &   $\overline{V}$ & $V_{gr}$  \\
Jun 2006    &  UT & mass &  $''$ & $''$ & m/s   &       m/s  \\
\hline
19/20 & 23:38--23:44  & 1.7 &1.38 & 0.63 &    30 & 5.6  \\
21/22 & 1:40--1:49    & 1.5 &1.21 & 0.68 &    13 & 2.5  \\ 
22/23 & 23:07--23:16  & 1.1 &0.54 & 0.27 &    16 & 6.4  \\ 
\hline
\end{tabular}
\end{table}

\begin{figure*}
\centerline{
\includegraphics[width=5.5cm]{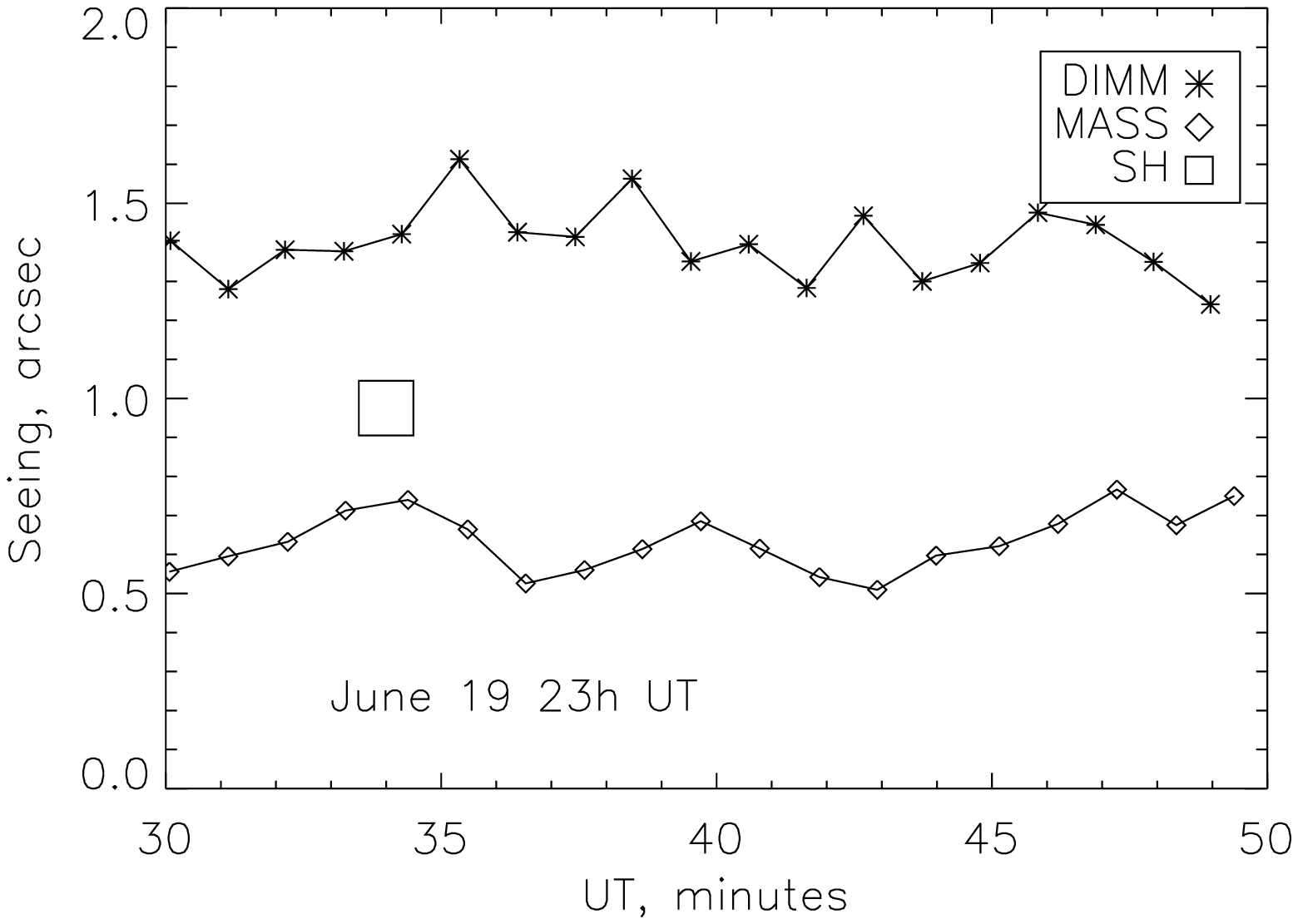}
\includegraphics[width=5.5cm]{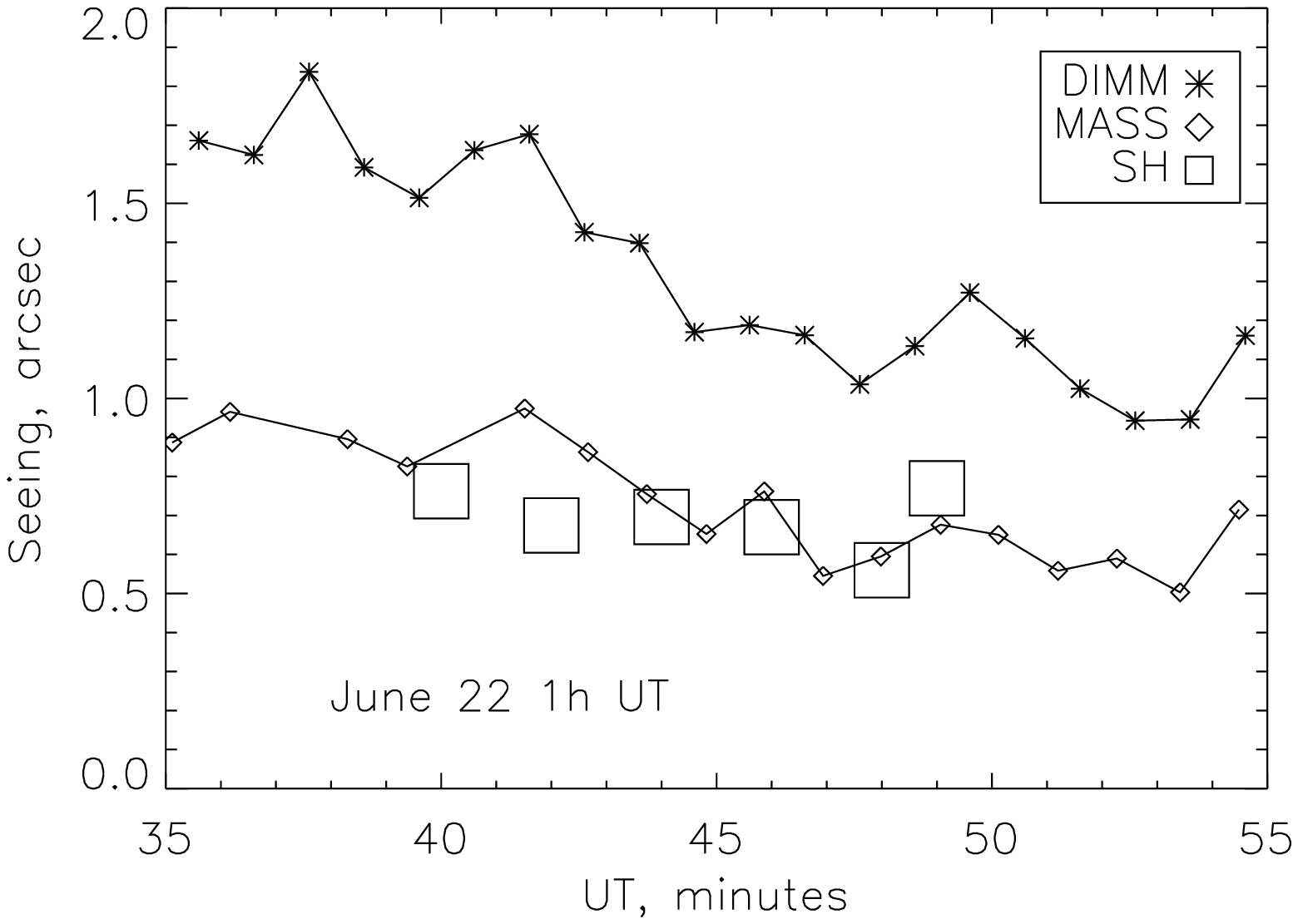}
\includegraphics[width=5.5cm]{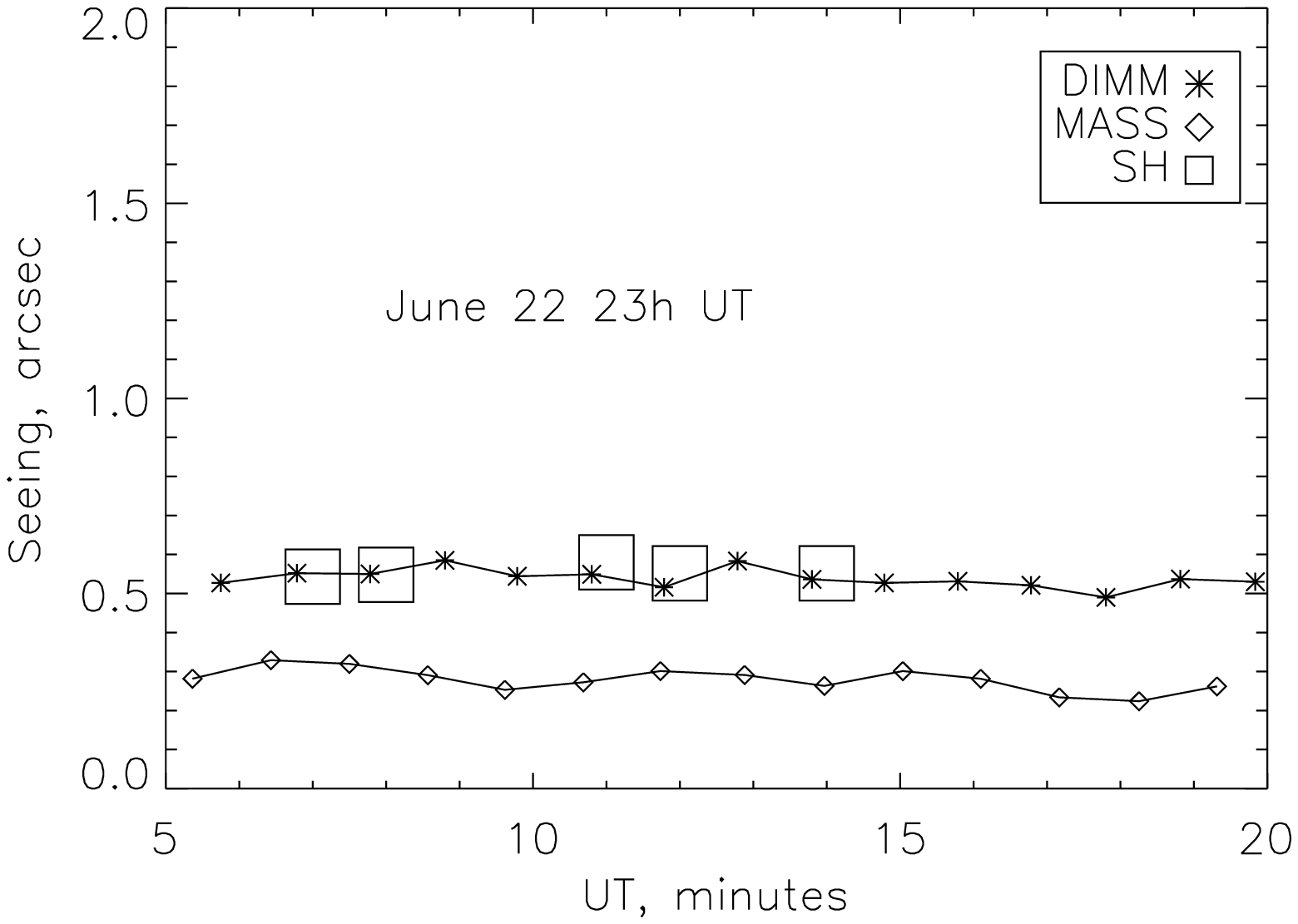}
}
\caption{Temporal evolution of the seeing $\epsilon$ and the seeing in
  the  free atmosphere $\epsilon_{FA}$  as measured  by DIMM  and MASS
  respectively during  the data acquisition  periods on 3  nights. The
  seeing  derived  from  the  SH   spots  and  reduced  to  zenith  is
  over-plotted as large squares.
\label{fig:seeing} }
\end{figure*}

The data  for this study have been  obtained by A.S.  on  June 19, 21,
and   22,  2006.    Table~\ref{tab:nights}   lists  relevant   average
parameters for  each data set. On  all nights the sky  was clear, with
stable   air   temperature   and   very  low   humidity.    Individual
(non-averaged)   data   from   DIMM    and   MASS   are   plotted   in
Fig.~\ref{fig:seeing}   to  characterize   the   variability  of   the
turbulence. The  conditions were rather  stable on all 3  nights, with
the DIMM  seeing always dominated  by the ground layer.   However, the
seeing  derived from  the SH  spots indicates  that the  ground seeing
contribution at VLT  was less than at DIMM  on June 19/20 and 21/22.

\subsection{VISIR data}

Mid-IR images of bright stars  were obtained with VISIR in two filters
called    PAH1    (8.6\,$\mu$m)    and    Q2    (18.7\,$\mu$m),    cf.
Table~\ref{tab:par}.  A standard  chopping-nodding technique was used.
At  each  telescope position  (nod),  the  image  was shifted  on  the
detector  back and  forth by  modulation (chop)  of the  VLT secondary
mirror M2  with a period of 2-4\,s.  The chop throw was  $10''$ in the
North-South  direction, with  a little  pause to  stabilize  the image
after each  chop.  For  the PAH1  filter, for example,  a total  of 30
images with 2\,s  exposure in each chop position  are taken to produce
the data cube with cumulative exposure of 60\,s. Then the telescope is
moved by $10''$ to the East and  a second data cube is taken.  Here we
do  not take advantage  of the  nodding and  analyze only  the average
difference between images in two chopping positions A and B.
 This   A$-$B  difference  suppresses  the  background   and  its  slow
drifts. It is  averaged over all image pairs in  the cube and contains
positive (A) and negative (B) images of the same star, considered here
separately as two independent realizations of the PSF.

\begin{figure}
\centerline{\includegraphics[width=8.5cm]{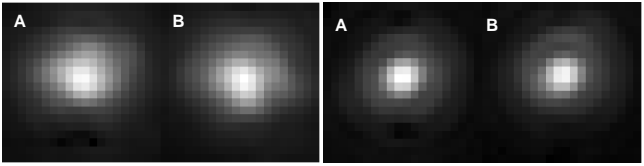}}
\caption{The pairs  of positive and  negative PSFs in the  PAH1 filter
registered  with VISIR on  June 21/22  (left, file  5) and  June 22/23
(right,  file 1).   Only the  central 20x20  pixels ($1.5''$)  of each
image are displayed with a square-root intensity stretch.
\label{fig:pah} }
\end{figure}

\begin{figure}
\centerline{ \includegraphics[width=8.5cm]{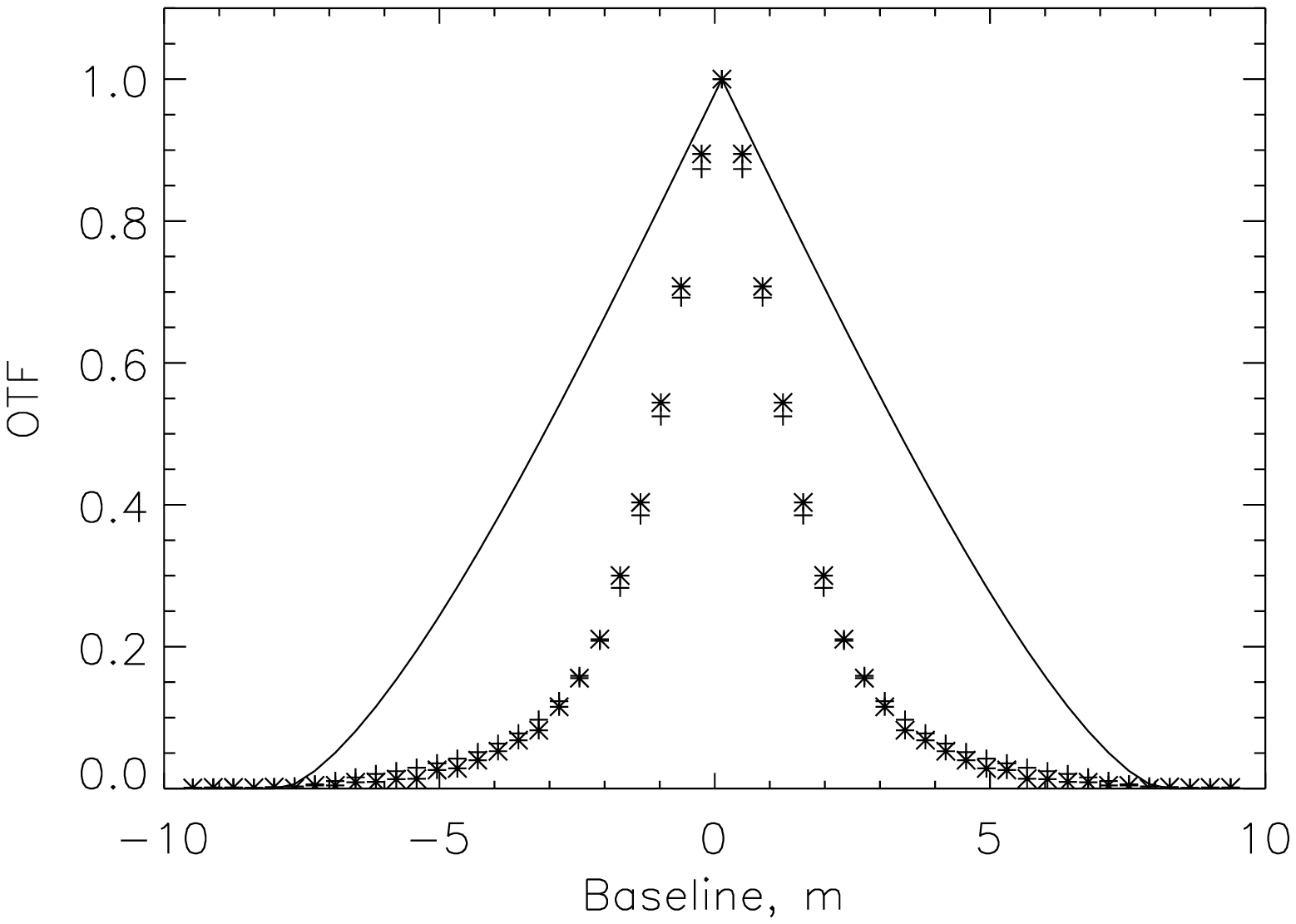} }
\centerline{\includegraphics[width=8.5cm]{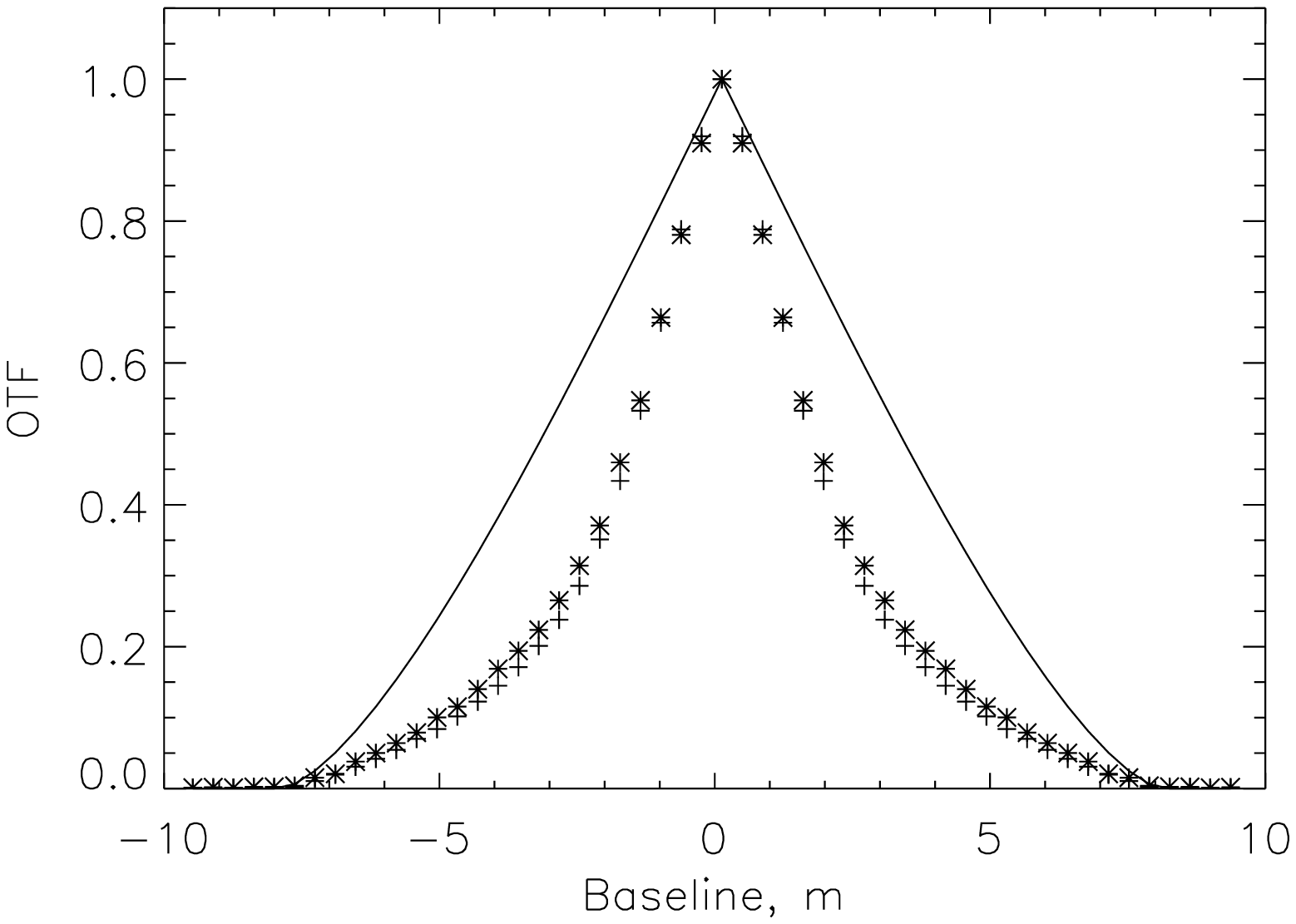}}
\caption{The cuts  of the OTFs along  $f_x$ axis for  the VISIR images
displayed  in Fig.~\ref{fig:pah} (top  -- June  21/22, bottom  -- June
22/23).  Asterisks -- positive images, plusses -- negative images. The
line shows the diffraction-limited OTF $T_0(\lambda f)$.
\label{fig:pahotf} }
\end{figure}

The  positive and  negative  PSFs  are extracted  as  two 64x64  pixel
($4.8''$)  subsections and  processed in  parallel. The  background is
computed  in the corners  of these  images (outside  the radius  of 32
pixels) as a median and subtracted. For reference only, these PSFs are
approximated with 2-dimensional Gaussians  to determine the Full Width
at Half Maximum  (FWHM) $\epsilon_l$ and $\epsilon_s$ in  the long and
short  axes.   The ellipticity  is  computed  as  $e =  (\epsilon_l  -
\epsilon_s)/(\epsilon_l + \epsilon_s)$ and the position angle $\theta$
of  the long  axis (counted  from the  $x$-axis  counter-clockwise) is
recorded as well.

\begin{figure}
\centerline{
\includegraphics[width=8.5cm]{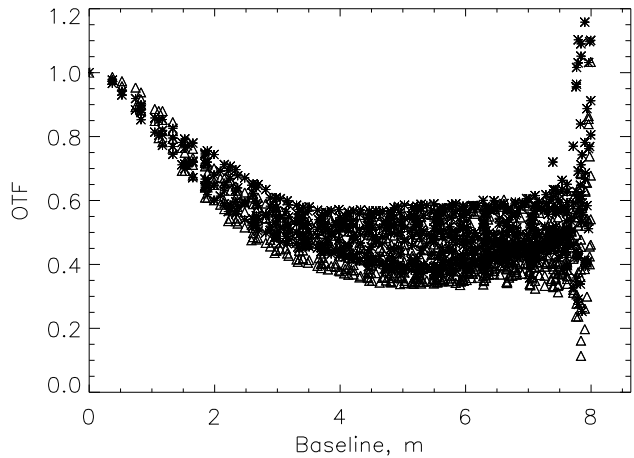}
}
\caption{The atmospheric  OTF in the VISIR PAH1  filter obtained after
  correcting the observed OTFs for diffraction on June 22/23, file 1. All
  points are  plotted without radial averaging (asterisks  -- image A,
  triangles -- image B) to show the asymmetry. Compare with
  Fig.~\ref{fig:pahotf}.   
\label{fig:aotf} }
\end{figure}

Figure~\ref{fig:pah} shows typical examples  of the individual PSFs in
the PAH1 filter on two nights.  On  June 21/22, the PSF is blurred by the
turbulence. It  is slightly elongated  ($e =  0.1...0.2$) and  has some
structure, different between positive and negative images. On June 22/23,
under a better seeing, the PSF is closer to a diffraction-limited one,
but still elongated ($e = 0.08...0.12$).  The elongation can be caused
by a combination of  several factors including some residual low-order
aberrations,  vibrations, etc. However,  the dominant  contribution to
the  image elongation  is likely  related to  the  tilt anisoplanatism
between the  guide star and the  object, estimated in  Appendix C. The
direction of the elongation points approximately to the guide star.

Each PSF is  Fourier-transformed and the normalized modulus  of the FT
is identified with the  observed $T(f)$ (Fig.~ \ref{fig:pahotf}).  The
spatial frequencies $f$  are translated to the baselines  $r = \lambda
f$.  A small correction is needed for the normalization (cf.  Appendix
B).  The experimental OTF $T(f)$ is divided by the calculated $T_0(f)$
to  obtain  the atmospheric  OTF  and  then  the SF.   In  calculating
$T_0(f)$, we  use the  pupil diameter $D=8.115$\,m  because it  is not
vignetted   by   the   cold   stop  inside   the   VISIR   instrument.
Figure~\ref{fig:aotf}  shows  the results  of  this  division for  the
sharpest PAH1 image  (compare with Fig.~\ref{fig:pahotf}, bottom). The
saturation of  the OTF (hence SF)  is obvious. The  vertical spread is
caused by the  asymmetry and the differences between  the images A and
B.

Note that at small baselines  the derived atmospheric OTFs and SFs are
sensitive to  the normalization errors,  while at large  baselines the
experimental $T(f)$  becomes noisy  and its modulus  is biased  by any
image defects such as noise and bad pixels.  For these reasons, we use
for further  analysis the  baseline range from  0.4\,m to  4\,m, where
$T(f)$ is most reliable.  In the following, the asymmetry is neglected
and  the atmospheric  SFs  are calculated  from the  radially-averaged
OTFs.

The  data in  the Q2  filter  are rather  noisy compared  to the  PAH1
filter.   The PSFs  are nearly  diffraction-limited.  At  some nodding
positions,  the  image  is  affected  by bad  detector  pixels  and/or
horizontal  stripes in the  background.  The  OTFs registered  on June
22/23  do not  differ  from the  diffraction-limited  ones within  the
errors.  Therefore,  no reliable estimates  of the atmospheric  SF are
derived from the  Q2 images on this night and we  can only affirm that
the phase fluctuations were much less than 1\,rad at 18.7\,$\mu$m.

\subsection{Shack-Hartmann data}

 Long-exposure  images  from  the  Shack-Hartmann (SH)  sensor  were
recorded quasi-simultaneously with the data. A typical exposure time is 45\,s,
but the SH ``sees'' the star only during 1/2 of the chopping cycle. On
the  other  hand,  the  guiding  was  done  continuously  because  the
guiding ``box'' moved to compensate for the chopping.

\begin{figure}
\centerline{\includegraphics[width=8cm]{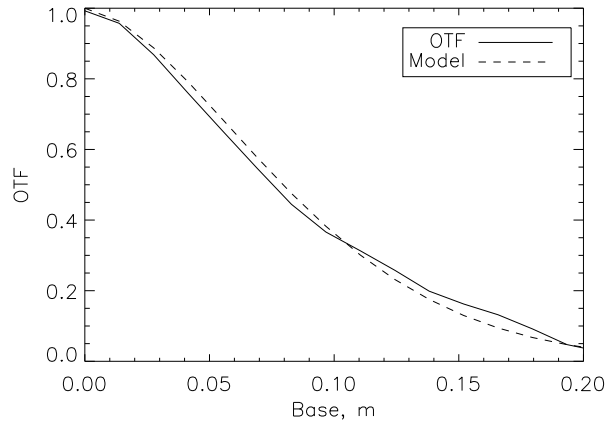} }
\caption{The  atmospheric  OTF  derived   from  the  average  SH  spot
de-convolved  by  the  average   reference spot (June 22/23, 23:07). 
The  fitted
Kolmogorov model is plotted in dashed line. 
\label{fig:Fried} }
\end{figure}

\begin{figure*}
\centerline{\includegraphics[width=16cm]{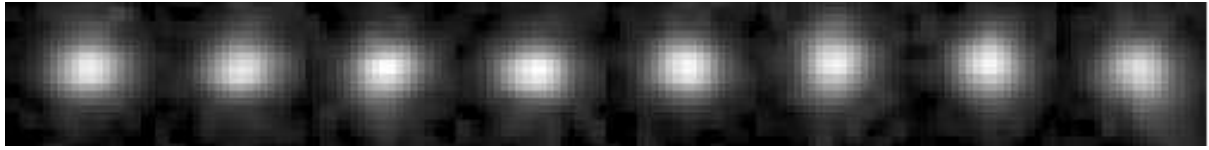} }
\caption{Mosaic of 8  individual good SH spots (June  22/23, 23:15), each
  in a  $3'' \times 3''$  field. The data  are taken at air  mass 1.1,
  with negligible atmospheric dispersion.  The ellipticity of
  the average spot is $e = 0.08$.
\label{fig:sh} }
\end{figure*}

The SH  images are ``raw'', i.e.  not corrected for  bias, hot pixels,
flat field, etc.  The geometry of the lenslet array is square, with 24
spots across pupil diameter,  hence sub-aperture size $d=0.34$\,m. The
pixel scale  $0.30''$ was calculated from the  opto-mechanical data of
the SH  sensor and telescope.  However, the real pixel  scale depends,
among other things,  on the actual distance between  the lenslet array
and  the detector.  So we  obtained images  of the  known  double star
HIP~73246  with a separation  of $3.512''$  measured by  Hipparcos and
derived the pixel scale $0.280'' \pm 0.004''$ for the SH sensor at the
Cassegrain focus of UT3.

The images  of individual spots  show various distortions and  are all
elongated  in  one  direction  due  to  the  un-corrected  atmospheric
dispersion  (except for the  data taken  near zenith).   Initially, we
selected for the analysis only  the sharpest spots in each frame.  The
spots  were extracted,  over-sampled, re-centered  and  averaged.  The
influence  of local  CCD defects  is  reduced by  the averaging.   The
modulus of the Fourier Transform  $T({\bf f})$ of the average spot was
calculated and  normalized so that  $T(0)=1$.  Then it was  divided by
the diffraction-limited OTF for the square aperture $T_0({\bf f}) = (1
- | \lambda f_x /d|) \; (1 -  | \lambda f_y /d|)$, assuming $\lambda =
0.6$\,$\mu$m.

We presume that the elongation of the spots is caused primarily by the
atmospheric  dispersion. The  long axis  of $T({\bf  f})$ is  found by
fitting a 2-dimensional Gaussian. The cut along this axis is fitted to
the Kolmogorov atmospheric OTF (eq.~\ref{eq:Dphi}), giving an estimate
of the Fried's parameter $r_{0, Kolm}$ and seeing $\epsilon_{SH} = 0.98
\lambda/r_{0, Kolm}$ (Fig.~\ref{fig:Fried}).  The atmospheric  SF at
short baselines is derived from the same cut.

We  found that  even the  sharpest  spots were  distorted by  residual
aberrations in the lenslets. A mosaic of selected sharp spots recorded
under  good  seeing  near  zenith  (Fig.~\ref{fig:sh})  shows  various
degrees of  distortion and elongation.   This became apparent  on June
22/23, under excellent conditions, when the SH measured a seeing of about
$0.75''$, worse  than the  DIMM seeing. To  overcome this  problem, we
used the image of the reference point source.  A total of 306 sharpest
reference spots  were selected, re-centered  and averaged in  the same
way as  the star images.  Then  we re-processed the  data by selecting
the same  spots and de-convolving  them by the average  reference spot
instead of $T_0({\bf  f})$.  A good agreement between  DIMM and SH was
reached (Fig.~\ref{fig:seeing}).

\section{Results}

\begin{table*}
\center
\caption{Data log}
\label{tab:log}
\medskip
\begin{tabular}{lll c   ccc  ccc  c c }
\hline
Time & File & Filt &  $\epsilon_{SH}$,  & \multicolumn{3}{c}{Image A} &  \multicolumn{3}{c}{Image B}   & $r_0$, & $L_0$,\\
UT   &      &      & $''$               & $\epsilon_s$ & $e$ & $\theta$ &$\epsilon_s$ & $e$ & $\theta$ &  m     & m     \\
\hline
 \multicolumn{3}{c}{June 19/20}                        \\
23:38 & 5 & Q2 &  1.348 & 0.573 &  0.09 &    22  &0.582 &  0.08 &    23 & 0.082 & 40 \\
23:40 & 6 & Q2 &   -    & 0.674 &  0.07 &    38  &0.663 &  0.08 &    43 & 0.086 & 200\\
23:42 & 7 & Q2 &   -    & 0.643 &  0.03 &    30  &0.624 &  0.03 &     9 & 0.086 & 200\\
23:44 & 8 & Q2 &   -    & 0.585 &  0.05 &    22  &0.578 &  0.08 &    17 & 0.082 & 50 \\
\multicolumn{3}{c}{June 21/22}  \\
1:40  & 1 & Q2  & 0.980 & 0.506 &  0.10 &   -25 &0.513 &  0.09  &  -30  & 0.112 & 50 \\
1:42  & 2 & Q2  & 0.867 & 0.512 &  0.08 &   -19 &0.530 &  0.08  &  -30  & 0.115 & 100\\
1:44  & 3 & Q2  & 0.895 & 0.515 &  0.08 &   -19 &0.529 &  0.06  &  -28  & 0.117 & 200\\
1:46  & 4 & Q2  & 0.861 & 0.493 &  0.03 &   -20 &0.505 &  0.02  &  -24  & 0.113 & 60 \\
1:48  & 5 & PAH1& 0.719 & 0.379 &  0.10 &    -1 &0.388 &  0.08  &   -9  & 0.110 & 30 \\
1:49  & 6 & PAH1& 0.989 & 0.373 &  0.17 &    -7 &0.400 &  0.19  &    3  & 0.111 & 40 \\
\multicolumn{3}{c}{June 22/23}  \\				  	 			   
23:07 & 1 & PAH1 & 0.583& 0.239 &  0.08 &    23 &0.247 &  0.09  &   21  & 0.181 & 35 \\
23:08 & 2 & PAH1 & 0.575& 0.245 &  0.12 &    32 &0.251 &  0.09  &   22  & 0.182 & 40:\\
23:11 & 3 & Q2   & 0.580& 0.457 &  0.02 &    14 &0.456 &  0.01  &   27  & 0.186 & 60:\\
23:12 & 4 & Q2   & 0.614& 0.455 &  0.02 &    20 &0.454 &  0.02  &    8  & 0.186 & 60:\\
23:14 & 5 & Q2   & 0.584& 0.448 &  0.04 &    20 &0.459 &  0.02  &   17  & 0.194 & 200:\\
23:16 & 6 & Q2   & 0.584& 0.451 &  0.02 &     4 &0.448 &  0.02  &   14  & 0.194 & 200:\\
\hline
\end{tabular}
\end{table*}

   The   results   of   image   processing   are   gathered   in   the
Table~\ref{tab:log}.  The  time (to 1~min)  refers to the end  of each
acquisition.  The  seeing at 0.6\,$\mu$m  estimated from the  SH spots
$\epsilon_{SH}$ is listed  as well (it is not reduced  to zenith as in
Fig.~\ref{fig:seeing}).  The  next columns give the  parameters of the
elliptical Gaussians  approximating the positive (A)  and negative (B)
mid-IR images: the FWHM  $\epsilon_s$ of the short axis  (in arcseconds), the
ellipticity $e$, and  the position angle $\theta$ of  the long axis in
degrees.

\begin{figure}
\centerline{\includegraphics[width=8cm]{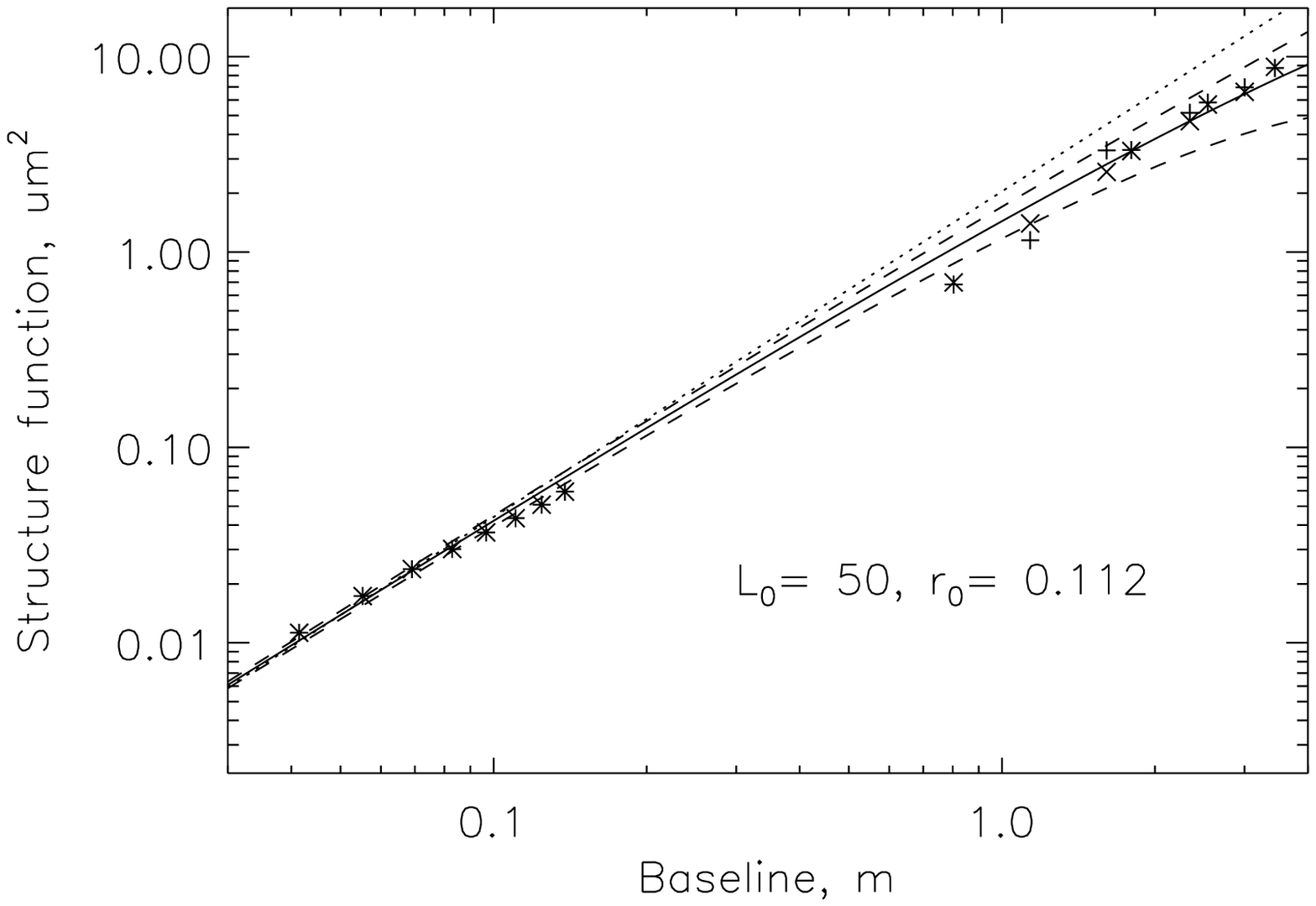}}
\centerline{\includegraphics[width=8cm]{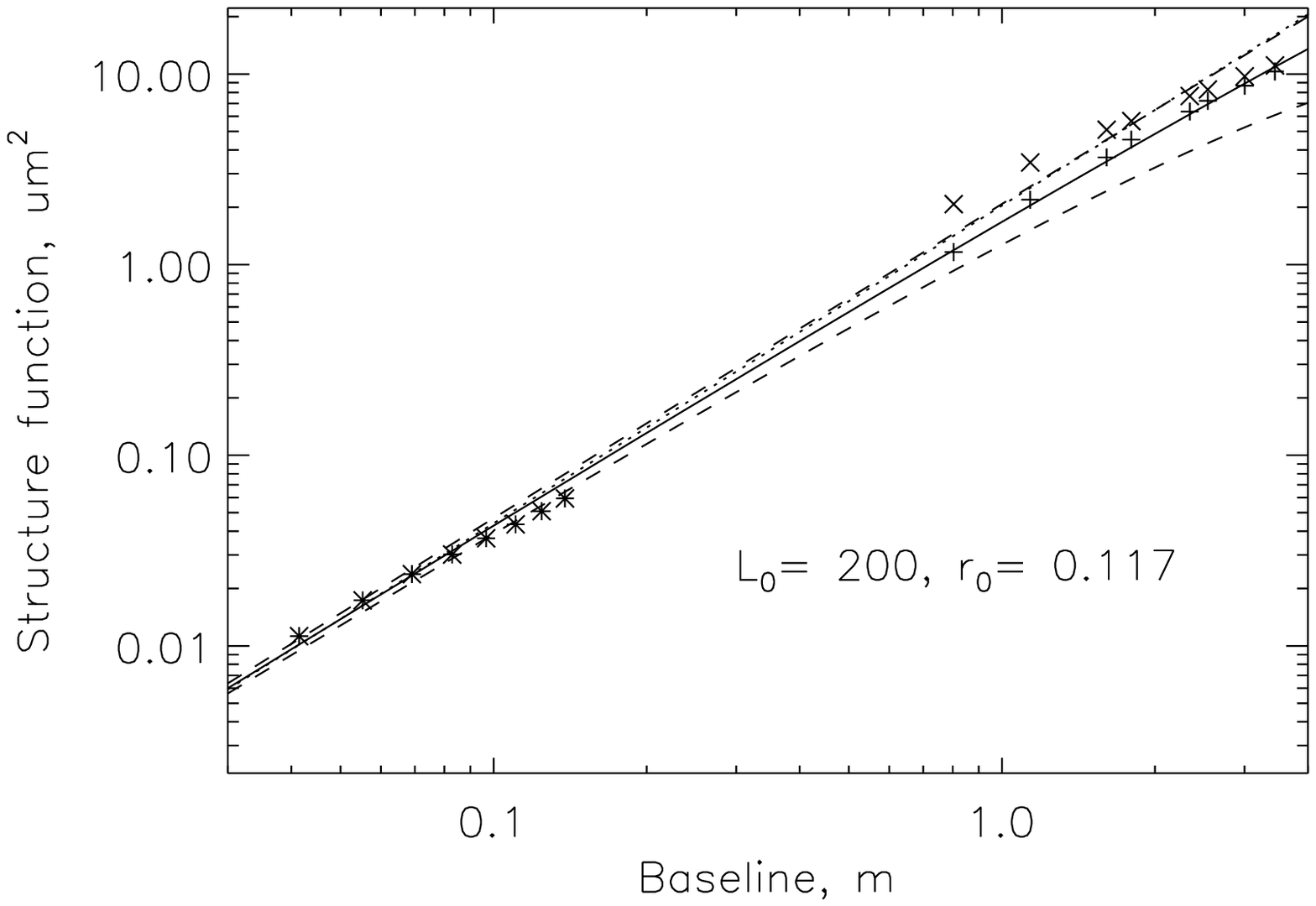}}
\caption{Structure functions  on June 21/22, Q2 filter  (files 1, top,
and 3,  bottom).  The SFs  derived from the  VISIR images A and  B are
plotted  as plusses  and crosses,  the  SFs from  the SH  spots --  as
asterisks.  Full lines  -- VK models, dashed lines  -- subtraction and
addition of $D^{tilt}$, dotted lines -- Kolmogorov SFs.
\label{fig:sf21-1} }
\end{figure}

\begin{figure}
\centerline{\includegraphics[width=8cm]{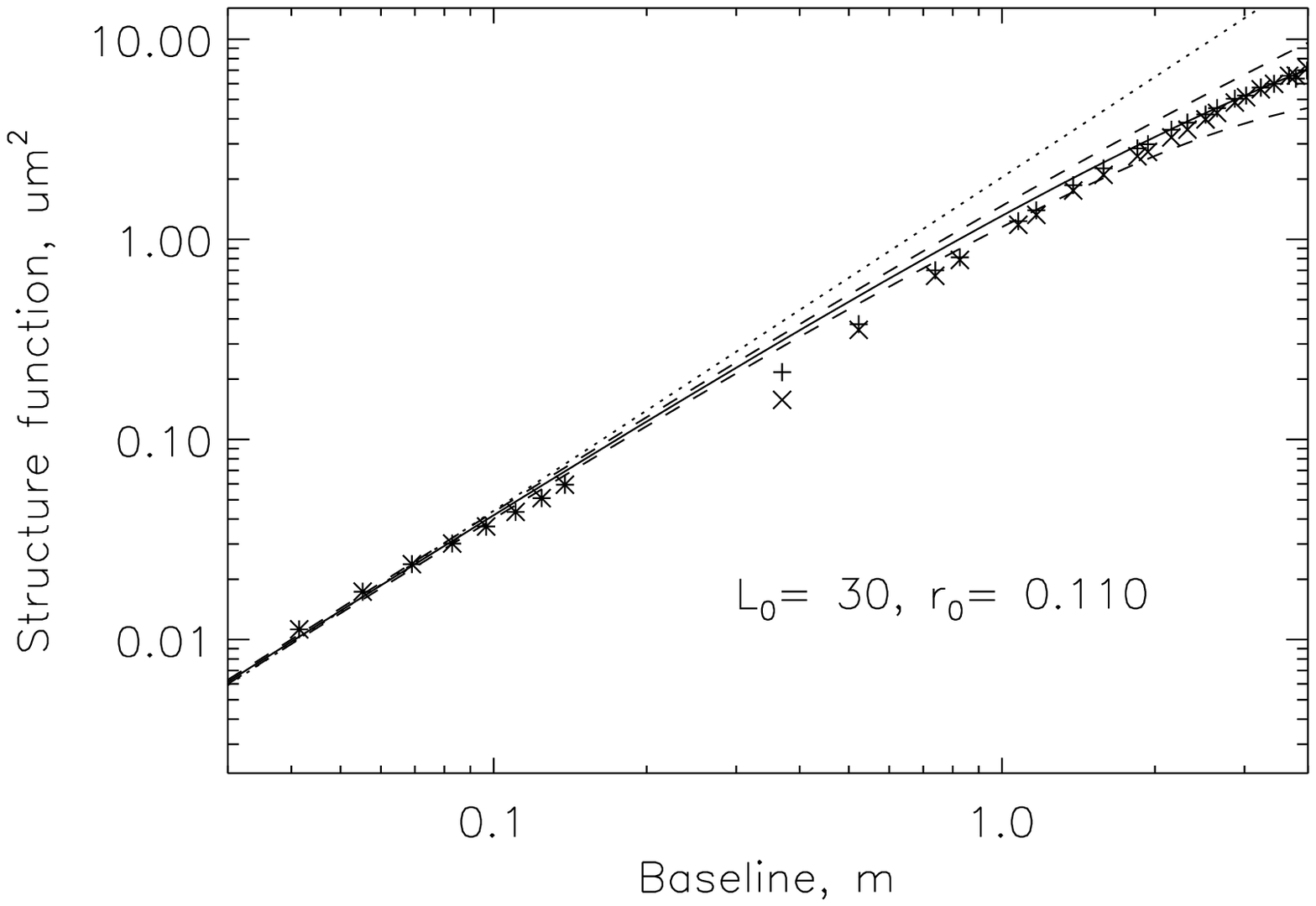}}
\centerline{\includegraphics[width=8cm]{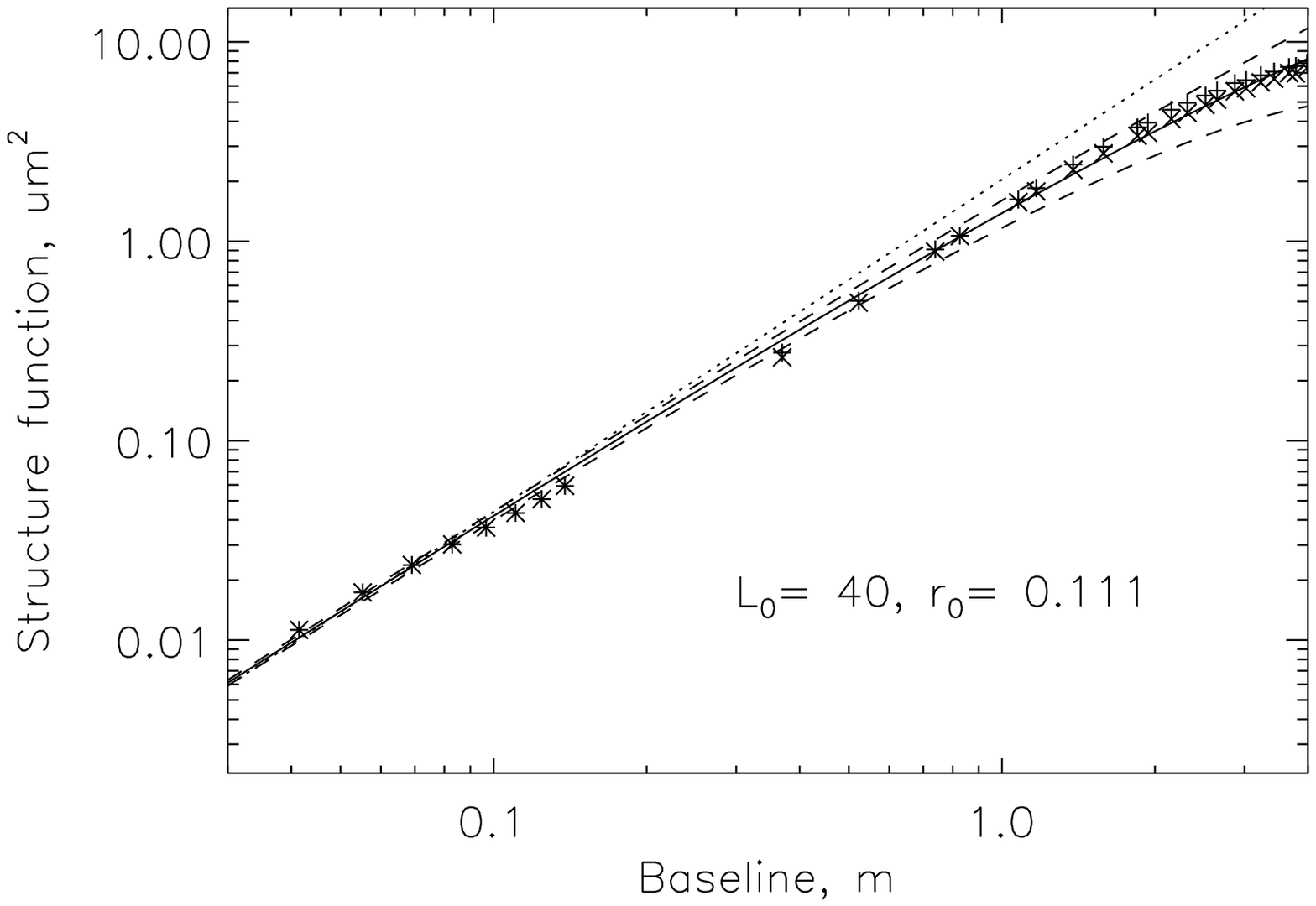}}
\caption{Structure functions  as in  Fig.~\ref{fig:sf21-1}  for  the PAH1  filter  (June
  21/22, files 5, top, and 6, bottom).
\label{fig:sf21-5} }
\end{figure}

\begin{figure}
\centerline{\includegraphics[width=8cm]{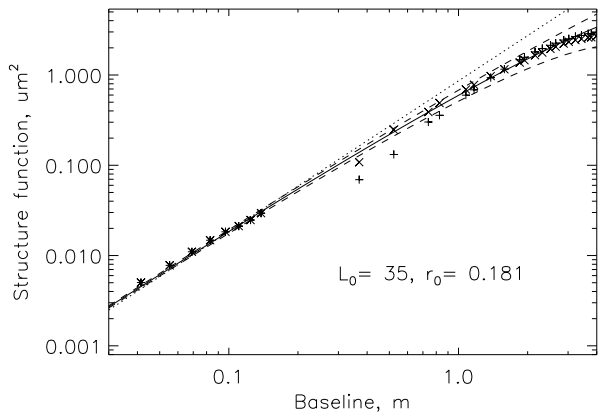}}
\caption{Structure  function  as  in Fig.~\ref{fig:sf21-1}  for  June
22/23, PAH1  filter, file 1. 
\label{fig:sf22} }
\end{figure}

The  SFs derived from  the visible  and mid-IR  PSFs are  converted to
linear units ($\mu$m$^2$) by multiplying them with $(\lambda/2 \pi)^2$
and         combined         on         the         same         plots
(Figs.~\ref{fig:sf21-1},\ref{fig:sf21-5},\ref{fig:sf22}).    They  are
compared to the VK models  (Appendix A).  The model parameters $r_0$
are derived from  the SH spots, and the outer  scale $L_0$ is selected
to match the data qualitatively. These parameters are also listed in
Table~\ref{tab:log}.

The exact  degree of tip-tilt compensation by  the field stabilization
servo  cannot be  evaluated.  A  large  part of  the atmospheric  tilt
produced by the  ground layer is compensated, but   tilts from high
layers are actually amplified  (Appendix C). However, the total effect
of the tilt compensation is not large and cannot explain the deviation
of  the SFs from  the Kolmogorov  model.  In  the figures,  the dashed
lines  show the  VK models  with complete  tilt compensation  and tilt
doubling, thus bracketing possible  effects of the field stabilization
system.

Figure~\ref{fig:sf21-1}   shows   the   data  from   two   consecutive
acquisitions made with an interval  of only 4\,min.  We see that $L_0$
increased from 50\,m  to 200\,m.  Further data show  that it decreased
again in  the next 6\,min.   (Fig.~\ref{fig:sf21-5}).  Such ``bursts''
of $L_0$ are typical \citep{GSM}.

On  June 22/23,  the  images in  the Q2  filter  are so  close to  the
diffraction limit that the SFs  derived from them are uncertain ($L_0$
values marked by colons in Table~\ref{tab:log}). The SFs derived from
the PAH1 image shows saturation (Fig.~\ref{fig:sf22}).

\section{Conclusions}

We were able to measure directly the structure function of atmospheric
wave-front distortions  at the VLT  up to metric scales.   The results
show a  broad agreement with the  VK turbulence model.   Hence, we can
use this model with an increased degree of confidence  for predicting
the   long-exposure   PSF   in   the  infrared   or   evaluating   the
deformable-mirror stroke.

Interpretation of the measured  SFs in terms of atmospheric turbulence
model   cannot  be  done   without  reservations,   however.   Several
instrumental effects bias these SFs.  Instead of uncertain modeling of
these effects, we simply present  the results ``as they are'' and hope
that  new,  deeper  studies will  be  prompted  by  this work.  It  is
preferable to  use a  good-quality optical imager  rather than  SH for
continuing this study.

The standard  theory predicts  an improvement of  the image size  in a
very  large telescope  (neglecting  diffraction) as  $\lambda^{-1/5}$,
e.g. by 1.70 times between  0.6\,$\mu$m and 8.6\,$\mu$m. In fact, on a
good  night  the  VLT  image  quality at  8.6\,$\mu$m  is  limited  by
diffraction, and  we observe a clear  saturation of the  SF.  It means
that the  $\lambda^{-1/5}$ scaling does  not work.  In a  larger, 30-m
telescope,  the  FWHM  resolution  at 8.6\,$\mu$m  will  be  $0.067''$
(diffraction-limited) for  the VK  model with decametric  outer scales
because the SF saturates at large baselines.  This example illustrates
a dramatic effect of turbulence model for predicting the long-exposure
image quality in the IR, demonstrated here  experimentally. 

\section*{Acknowledgments}

We thank  Stephane Guisard  for obtaining reference  SH images  on our
request. A suggestion by anonymous Referee to de-convolve the SH spots
with images of the reference source helped us to resolve the problem
of lenslet aberrations.

\appendix

\section{Models of turbulence}

The  von  K\'arm\'an  (VK) turbulence  model describes  the  spatial  power
spectrum   of   atmospheric    wave-front   phase   by   the   formula
\citep{Tatarsky61,Sasiela94,GSM}
\begin{equation}
W_\varphi (\kappa) = 0.0229 r_0^{-5/3} (\kappa^2 + L_0^{-2})^{-11/6} ,
\label{eq:VK}
\end{equation}
where  $\kappa$ is  the modulus  of  the spatial  frequency (one  over
period).   The model has  two parameters,  the coherence  radius $r_0$
(also called Fried  radius) and the outer scale  $L_0$. The Kolmogorov
turbulence  model is  a specific  case  of (\ref{eq:VK})  with $L_0  =
\infty$. 

The phase  structure function  (SF) $D_{\varphi}(r)$ is  obtained from
$W_\varphi (\kappa)$.   An analytic  expression for the  SF in  the VK
model can be found in \citep{Tatarsky61} and in \citep{Tok02}.  In the
limit $L_0 = \infty$ (Kolmogorov model) the SF is
\begin{equation}
 D_{\varphi}(r) =  6.8839 (r/r_0)^{5/3} .
\label{eq:Dphi}
\end{equation}

If we try to approximate a VK SF (or PSF) with a simpler Kolmogorov model, the
derived $r_{0,Kolm}$ will be larger than the true $r_0$. An
approximate relation between these parameters can be established
numerically. Here we use a formula adapted from \citep{Tok02},  
\begin{equation}
r_0 \approx r_{0,Kolm} \; \sqrt{ 1 - 1.5 (r_{0,Kolm}/L_0)^{0.356} } .
\label{eq:r0}
\end{equation}
  Thus we translate the $r_{0,Kolm}$  obtained by fitting the SH spots
to  the $r_0$  parameter  appropriate for  the  VK model  by means  of
(\ref{eq:r0}). This ensures a good match between the Kolmogorov and VK
SFs at short baselines (e.g. Fig.~\ref{fig:sf21-1}).

\section{Correction for the missing flux}

The wings of  the PSF (essentially caused by  diffraction) outside the
selected  field are cut  off, hence  $T(0)$ (integral  of the  PSF) is
under-estimated.  The fraction of energy  in the Airy PSF contained in
the circle of angular radius $a$ is \citep{BW}:

\begin{equation}
E(a) = 1 - J_0^2(\pi a d/\lambda) - J_1^2(\pi a d/\lambda) .
\label{eq:E}
\end{equation}
We take $a=  p N_{grid}=2.4''$, where $p = 0.075''$  is the pixel size
and $N_{grid}=32$ is the half-size of the PSF frame.

A second correction  is needed because we calculate  the background as
the  average   intensity  at  the   distance  $\approx  a$   from  the
center. Together with the background, we subtract some fraction of the
PSF.  This fraction  $\Delta E$,  integrated  over the  field, can  be
estimated by differentiating (\ref{eq:E}) as

\begin{equation}
\Delta E \approx \frac{a}{2 \Delta a} [E(a+\Delta a) - E(a) ] .
\label{eq:dE}
\end{equation}
We  take $\Delta  a =  \lambda/D$ to  average out  the  ``wiggles'' of
$E(a)$. For  PAH1 images,  $E(a) \approx 0.98$  and $\Delta  E \approx
0.01$. We divide $T(0)$ by $E(a) - \Delta E$ before the normalization,
thus accounting for the missing  flux. The correction in the Q2 filter
is  larger, reaching  5\%. Without  such correction,  we over-estimate
$T(f)$ at small baselines and  hence under-estimate the SF.  We see in
Fig.~\ref{fig:sf21-5} that  the first points  of mid-IR SFs  are still
below  the  model curves,  indicating  that  a  larger correction  for
missing flux was probably needed for the PAH1 images.

\section{Effect of the tip-tilt servo on the  SF}

The  variance   of  the  image  centroid  motion   in  one  coordinate
$\sigma^2_\alpha$  (in square radians)  can be  computed by  the known
formula
\begin{equation}
\sigma^2_\alpha = K_{tilt} (\lambda/D)^2 (D/r_0)^{5/3} , 
\label{eq:siga}
\end{equation}
where  $  K_{tilt}  =  0.170$  for  the  Kolmogorov  turbulence,  e.g.
\citep{Sasiela94}.  The image  motion is  achromatic because  the right
hand of  Eq.~\ref{eq:siga}  does not depend on  $\lambda$. For
the VK model, our numerical calculation leads to the approximation
\begin{equation}
\log_{10} K_{tilt} \approx -2.672 + 2.308 x - 0.898 x^2 ,
\label{eq:Ktilt}
\end{equation}
where $x = \log_{10} (L_0/D)$. The approximation is 
valid for  $L_0/D < 300$  with an accuracy  of better than 1\%.  For a
typical situation  at VLT, $L_0 =  20$\,m, $K_{tilt} =  0.013$, i.e. an
order of magnitude smaller than for the Kolmogorov model. 

The field stabilization servo measures the tilt of the guide star
$\alpha_2$, filters it with the closed-loop response $h(t)$ and
applies to compensate for the object tilt $\alpha_1$. The residual
tilt error is then  
\begin{equation}
\Delta \alpha = \alpha_1 - \alpha_2 \odot h .
\label{eq:da}
\end{equation}
It follows that the power spectrum  of the residual tilt variance is
\begin{equation}
W_{\Delta \alpha} (\nu) = W_\alpha(\nu) [ 1 + | \tilde{h}(\nu) |^2 ] 
- 2 {\rm Re} [ W_{12} (\nu)  \tilde{h}^*(\nu) ] ,
\label{eq:W1}
\end{equation}
where $ W_\alpha(\nu)$ is the power spectrum of the tilt, 
 $W_{12} (\nu)$ is the cross-power spectrum between guide star and
 object, and Re stands for the real part. The residual tilt variance
$\sigma^2_{\Delta \alpha}$ can be conveniently expressed as a fraction
 $r$ of the un-corrected variance,  
\begin{equation}
\sigma^2_{\Delta \alpha} = \int_0^\infty  W_{\Delta \alpha} (\nu) 
{\rm d} \nu = r \sigma^2_\alpha .
\label{eq:r}
\end{equation}
The servo response is
 modeled here by a simple integrator with a 3-db cutoff frequency
 $\nu_0$, 
\begin{equation}
\tilde{h} (\nu) = \frac{ 1} { 1 + i (\nu/\nu_0) } .
\label{eq:h}
\end{equation}

\begin{figure}
\centerline{\includegraphics[width=8cm]{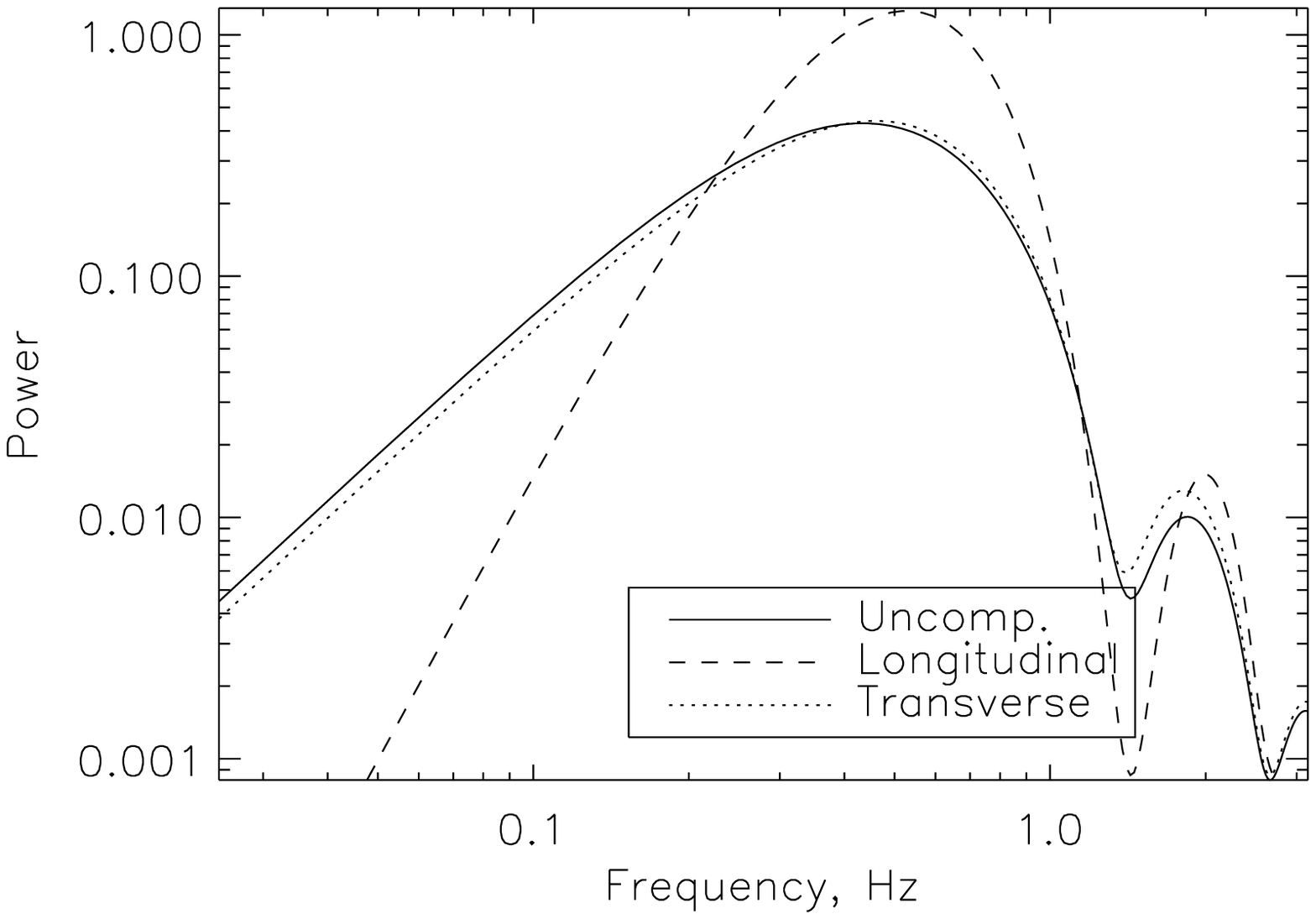} }
\caption{Temporal power  spectra of the tilt (full  line) and residual
  tilts in  longitudinal and transverse  directions after compensation
  with  a 1-Hz  servo. Parameters:  $D=8.115$\,m, $L_0  = 20$\,m,  $V =
  10$\,m/s, $s = 0.72D$, $\eta = 0$, $\nu_0 = 1$\,Hz.
\label{fig:cross} }
\end{figure}

The power spectrum and cross-spectrum  of tilt for VK turbulence model
is computed by Avila et al. (1997). For a single layer moving with the
speed  $V$ at direction  $\eta$ ($\eta=0$  for the  wind blowing  from the
object  to  the  guide   star),  the  cross-spectrum  depends  on  the
separation $s$ of  the beam footprints.  For example, a  layer at $H =
4$\,km and a guide star at $\theta  = 5'$ from the object lead to $s =
\theta  H =  5.8$\,m.  In  these conditions,  the 1-Hz  servo actually
increases the tilt variance:  $\sigma^2_{\Delta \alpha} = [2.36, 1.03]
\times  \sigma^2_\alpha $ for  $V =  10$\,m/s, where  the coefficients
$[r_l, r_t] =  [2.36, 1.03]$ refer to the  longitudinal and transverse
directions.  This situation is illustrated in Fig.~\ref{fig:cross}. On
the other  hand, for  the ground layer  ($s=0$) and $V=  5.5$\,m/s, we
obtain $r_l = r_t \approx 0.1$, i.e.  a good tilt correction.

Residual  tilt errors  of the  VLT field  stabilization system  can be
modeled. However, the input information  for such a model must include
the closed-loop servo  response $\tilde{h}(\nu)$, altitude profiles of
turbulence  $C_n^2(h)$ and $L_0(h)$,  profiles of  the wind  speed and
direction, and the  geometry of the object and guide  star.  We do not
have all this information for  the data at hand.  Instead, we evaluate
the effects of  the servo by assuming either $r_l =  r_t = 0$ (perfect
compensation) or $r_l = r_t = 2$ (un-correlated tilts). The first case
is closer to reality when a large part of turbulence is near the
ground. 

The wave-front  structure function $D(x,y)$ can be  represented by the
combination of the quadratic part caused by tilts $ D^{tilt}(x,y)$ and
the tilt-removed part $D^0$,
\begin{equation}
D(x,y) = D^0(x,y) + D^{tilt}(x,y) = D^0(x,y) + \sigma^2_\alpha (x^2 + y^2).  
\label{eq:D}
\end{equation}

When the field-stabilization servo is  at work, the resulting SF $D^g$
is
\begin{equation}
\begin{array}{ll}
D^g(x,y) & =  D^0(x,y) + \sigma^2_\alpha (r_l x^2 + r_t y^2) \\
& = D(x,y) + \sigma^2_\alpha(  r_l - 1) x^2   
+ \sigma^2_\alpha ( r_t - 1) y^2 ,  \\
\end{array} 
\label{eq:Dg}
\end{equation}
where  it  is assumed  that  the $x$-axis  is  directed  to the  guide
star. With  such orientation  of the coordinates,  we can  neglect the
cross-term proportional to $xy$,  which otherwise would be required in
the Eq.~\ref{eq:Dg}.   Our two options  (complete compensation, $r=0$,
or tilt doubling, $r=2$) correspond  to the subtraction or addition of
$D^{tilt}(x,y)$ from the VK model. The tilt variance $\sigma_\alpha^2$
is calculated with eqs.~\ref{eq:siga},\ref{eq:Ktilt}.

As an  example, consider the  case of file  5 (PAH1) on June  21/22. A
good match between MASS and SH seeing (Fig.~\ref{fig:seeing}) and MASS
profiles indicate  that most  of turbulence was  located in  a strong
layer at 4\,km.  Our model leads to $[r_l, r_t] = [2.4, 1.0]$ (assumed
parameters: $s=5.8$\,m, $V=10$\,m/s, $\eta = 0$, $L_0 = 30$\,m, seeing
$1.0''$,  $\nu_0   =  1$\,Hz).   We   compute  the  SF   according  to
Eq.~\ref{eq:Dg} and translate it to the PSF using Eq.~\ref{eq:T}.  The
resulting PSF  at 8.6\,$\mu$m  has a minimum  FWHM of $0.37''$  and an
ellipticity  of 0.08,  elongated towards  the guide  star  at position
angle  $-  39^\circ$.   The  actual  FWHM was  $0.38''$  with  $e=0.1$
elongated at $- 5^\circ$.  Thus, tilt anisoplamatism can qualitatively
explain the image elongation.

\bsp

\label{lastpage}

\end{document}